\documentclass[amsmath,twocolumn,superscriptaddress,prb,aps]{revtex4-1}
\usepackage{amsthm,amsfonts,graphicx,verbatim, color}
\usepackage{graphicx}
\usepackage{bm}
\usepackage[utf8]{inputenc}
\let\oldmarginpar\marginpar
\renewcommand\marginpar[1]{\-\oldmarginpar[\raggedleft\footnotesize #1]%
{\raggedright\footnotesize #1}}

\newcommand{\be}{\begin{equation}}
\newcommand{\ee}{\end{equation}}
\newcommand{\bea}{\begin{eqnarray}}
\newcommand{\eea}{\end{eqnarray}}

\renewcommand{\epsilon}{\varepsilon}
\renewcommand{\vec}[1]{{\bf #1}}

\def\beq{\begin{equation}}
\def\eeq{\end{equation}}
\def\bea{\begin{eqnarray}}
\def\eea{\end{eqnarray}}

\begin{document}

\title{Many body localization and delocalization in the two dimensional continuum}
\author{Rahul Nandkishore}
\affiliation{Princeton Center for Theoretical Science, Princeton University, Princeton, New Jersey 08544, USA}
\begin{abstract}
We discuss whether localization in the two dimensional continuum can be stable in the presence of short range interactions. We conclude that, for an impurity model of disorder, if the system is prepared below a critical temperature $T < T_c$, then perturbation theory about the localized phase converges almost everywhere. As a result, the system is at least asymptotically localized, and perhaps even truly many body localized, depending on how certain rare regions behave. Meanwhile, for $T > T_c$, perturbation theory fails to converge, which we interpret as interaction mediated delocalization. We calculate the boundary of the region of perturbative stability of localization in the interaction strength - temperature plane. We also discuss the behavior in a speckle disorder (relevant for cold atoms experiments) and conclude that perturbation theory about the non-interacting phase diverges for arbitrarily weak interactions with speckle disorder, suggesting that many body localization in the two dimensional continuum cannot survive away from the impurity limit. 
\end{abstract}
\maketitle

The field of localization in well isolated quantum systems originated in pioneering work by P.W. Anderson in 1958 \cite{pwa}. While it was rapidly established that non-interacting quantum particles in disordered potentials undergo localization for arbitrarily weak disorder in one and two dimensions \cite{LeeRMP}, and for strong enough disorder in three dimensions, the fate of interacting many particle systems in disordered environments has long been an open problem \cite{Fleishman}. Recent progress (\onlinecite{agkl, Gornyi} and especially \onlinecite{BAA}) has established that many body systems can also undergo localization for for sufficiently strong disorder - a phenomenon that has come to be known as many body localization (MBL). Recent years have seen an intense surge of interest in the MBL phenomenon (\cite{Oganesyan, pal, Znid} and many more - for a review see Ref.\onlinecite{arcmp}). MBL systems display an extensive suite of properties, including an emergent integrability \cite{Abanin, Lbits}, spectral functions of local operators that appear gapped at all temperatures \cite{ngh, johri} and exotic forms of quantum order that persist to high temperatures \cite{LPQO, Vosk, Pekker, Bauer, Bahri, lspt}. As a result, MBL represents an exciting new frontier for quantum statistical mechanics. 

However, common to almost all existing studies of MBL is an assumption that we are dealing with lattice systems. Lattice systems have the property that in the non-interacting limit, the single particle localization lengths are bounded, allowing for the development of a perturbation theory \cite{BAA}, the convergence of which guarantees at least the perturbative stability of MBL. Whether continuum systems with disorder can display MBL is a question that has largely been ignored, despite its importance as a fundamental question, its relevance for cold atoms experiments, and its implications for the burgeoning field examining localization in systems with translation invariant Hamiltonians \cite{grover, deroeck, Muller, glass}. In continuum systems, the localization length is unbounded, making it difficult to develop a perturbation theory. Recently, it has been argued \cite{Aleiner} that one dimensional systems without a periodic potential {\it can} exhibit MBL for sufficiently weak disorder, but the behavior in higher dimensions has not been established. Two dimensional systems in a continuum potential present a particularly interesting challenge, because the localization length is not only unbounded, but increases exponentially with energy at high energies. Can MBL occur in two dimensional continuum systems? 

In this paper, we examine the perturbative stability of localization in two dimensional disordered systems without a periodic potential. We conclude that in the impurity limit, when the two point correlation function of the disorder is independent of momentum transfer, perturbation theory about the localized phase converges almost everywhere, if the system is prepared below a critical temperature $T_c$ (which we calculate). As a result, the system is at least asymptotically localized \cite{deroeck}, and perhaps even many body localized, depending on how certain rare regions behave.  Above the critical temperature a perturbation theory about the non-interacting localized phase diverges for arbitrarily weak interactions - a divergence that may signify many body delocalization \cite{RNACP}. We calculate the boundary of perturbative stability of the localized phase in the interaction strength - temperature plane, and contrast the behavior with that of one dimensional systems, where a regime of perturbative stability persists to arbitrarily high temperatures. We also discuss the behavior away from the impurity limit. We conclude that for more general models of disorder (including the experimentally relevant case of the speckle potential), there is no regime of perturbative stability. Thus, for general models of disorder, there is may be no MBL phase (asymptotic or otherwise) in the two dimensional continuum - only a crossover to an almost localized regime. However, as the disorder distribution approaches the impurity limit, the crossover sharpens and approaches a true phase transition. 

This paper is structured as follows: we start by setting up the basic problem. We then analyze the problem within a certain approximation (assuming the disorder potential is in the impurity limit). We present the phase diagram, and discuss some unique properties of the localized phase contained therein. In particular, we argue that the localized phase presents an example of `marginal localization' as discussed in \cite{RNACP}. Next, we move beyond the impurity model of disorder, and consider more general disorder distributions (such as speckle potentials). We argue that the perturbation theory does not converge for such models, and that they likely undergo many body delocalization. In Appendix A we discuss various subtleties regarding the perturbation theory. In Appendix B we discuss various subtleties involving rare region effects (which are also briefly discussed in the main text). 

\section{The problem}

Consider a system of particles moving in the two dimensional continuum (with a random potential) and interacting via short range interactions. The (second quantized) Hamiltonian takes the form
\begin{widetext}
\begin{equation}
H = \int d^2 r \frac{\hbar^2}{2m} \nabla \psi^{\dag}_{\vec{r}} \nabla \psi_{\vec{r}}  + V(\vec{r}) \psi^{\dag}_{\vec{r}}\psi_{\vec{r}}+ \int d^2r d^2 r' g(|\vec{r}-\vec{r'}|) \psi^{\dag}_{\vec{r}} \psi_{\vec{r}} \psi^{\dag}_{\vec{r'}} \psi_{\vec{r'}}
\end{equation}
\end{widetext}
The argument may be constructed for bosons or fermions, whichever is more convenient. We choose to work with fermions for specificity. Fermions have the added advantage that the interaction strength can be tuned in cold atoms realizations via a Feshbach resonance. We choose to work with repulsive interactions, to eliminate the possibility of Cooper pairing. The interaction should be short ranged. For maximum simplicity, we assume an interaction that is a delta function in real space $g(\vec{r}-\vec{r'}) = g \delta(\vec{r}-\vec{r'})$. Finally, the potential $V(\vec{r})$ incorporates both the disorder, and also e.g. any trapping potential that may be present in  a cold atoms experiment. An important quantity in the following analysis will be the two point correlation function of the disorder, in Fourier space, defined as 
\begin{equation}
\gamma(k) = \int d^2r e^{i \vec{k} \cdot \vec{r}} \overline {V(\vec{0}) V(\vec{r})}
\end{equation}
where the over line indicates ensemble average over disorder, and where we have assumed that the disorder distribution is isotropic. We expect that $\gamma(0) \sim W^2 a^2$, where $W$ is the rms disorder potential, and $a$ is the disorder correlation length. 

We consider a system at a density $n$, prepared in equilibrium with a bath at a temperature $T$ so that a single particle orbital at an energy $E$ is occupied with a probability $\sim \frac{1}{1+\exp((E-\mu)/k_BT)}$.  The arguments presented below may also be generalized to arbitrary initial distributions (we will indicate how), but it is convenient to start with a Fermi-Dirac distribution. The principal strategy employed in the present work will be to start with the non-interacting limit $g \rightarrow 0$, where all states are localized, and then ask whether perturbation theory in small but non-zero interaction strength $g$ converges. If perturbation theory converges, then the interacting problem is also expected to be (many body) localized, whereas if perturbation theory diverges, then the analysis is indeterminate, and interaction mediated delocalization is a possibility. 

In order to discuss the perturbation theory, it is useful to first discuss the various length scales present in the problem. In addition to the disorder correlation length $a$, there is also the de-Broglie wavelength of the particles, equal to 
\begin{equation}
\lambda(E) = \frac{\hbar}{\sqrt{2mE}},
\end{equation} 
and a mean free path $l$. In the regime $l/\lambda  \ll 1$, all states are strongly localized with localization lengths of order $\lambda$, and the analysis of Ref.\onlinecite{BAA} establishes that perturbation theory about the localized phase converges, so that many body localization can occur. The theoretically unsolved regime is the regime $l/\lambda  \gg 1$. We note in particular that particles at the highest energies are inevitably in this regime, so that in order to understand whether many body localization can occur in the two dimensional continuum, we must understand whether localization can survive non-zero $g$ in the regime $l/\lambda \gg 1$.  

In the regime $l/\lambda \gg 1$, all states are localized in the limit $g \rightarrow 0$, but with localization length \cite{LeeRMP}
\begin{equation}
\xi = l \exp(\pi^2 l/\lambda).
\end{equation}
The wavelength is given by (1). Meanwhile, the mean free path $l$ may be estimated within a self consistent Born approximation (SCBA), which is valid when $l/\lambda \gg 1$, and which predicts
\begin{equation}
l = \frac{\hbar^3 \sqrt{2E/m} }{m \pi^2 \gamma(E)} = \frac{1}{\lambda} \frac{\hbar^4}{m^2 \pi^2 \gamma(E)}
\end{equation}
where we have allowed for the possibility that the two point function of the disorder might itself depend on energy. Substituting into the expression for localization length, we obtain
\begin{equation}
\xi(E) = \frac{\hbar^3 \sqrt{2E/m} }{m \pi^2 W^2 a^2}  \exp \left(\frac{\hbar^2 E}{m \pi \gamma(E)} \right).\label{xi}
\end{equation}
We note that the localization length is unbounded, and can grows exponentially large as we go to high energies. Whether this type of localization can survive non-zero interactions is the question we are interested in answering

\section{Impurity limit}
We begin by examining the impurity limit, in which the disorder correlation length is taken to be much shorter than all other length scales in the problem, and where we take $\gamma \sim W^2 a^2$ to be independent of energy. Since the correlation length of the disorder must be much smaller than all other length scales in the problem to be in the impurity limit, the limit we are considering is $a \rightarrow 0$ at constant $Wa$. In this impurity limit we get 
\begin{equation}
l = \frac{\hbar^3 \sqrt{2E/m} }{m \pi^2 W^2a^2 } = \frac{1}{\lambda} \frac{\hbar^4}{m^2 \pi^2 W^2a^2}
\end{equation}
Introducing $\lambda_W = \hbar/\sqrt{2mW}$ to be the de Broglie wavelength of a free particle with energy $W$, we can rewrite this as 
\begin{equation}
l = \frac{4 \lambda_W^4}{\pi^2 a^2 \lambda} \Rightarrow \frac{l}{\lambda} = \frac{4 \lambda_W^4}{\pi^2 \lambda^2 a^2}
\end{equation}
the condition for being in the `interesting' regime $l/\lambda \gg 1$ is 
\begin{equation}
\frac{2 \lambda^2_W}{\pi \lambda(E) a } \gg 1
\end{equation}
In this interesting regime, substituting Eq.(5) into Eq.(3), we find that the localization length depends on energy according to 
\begin{equation}
\xi(E) = \frac{\hbar^3 \sqrt{2E/m} }{m \pi^2 W^2 a^2}  \exp \left(\frac{\hbar^2 E}{m \pi W^2 a^2} \right).\label{xi}
\end{equation}

\subsection{Stability of localization}
We now ask whether localization is stable to the introduction of weak short range interactions. We do this in the manner of BAA: we compare the matrix element to the accessible level spacing. 

Let us consider the effect of interactions at a typical point in space, where two particles in states $\alpha$ and $\beta$ may hop (via the interaction) to states $\gamma$ and $\delta$. The matrix elements are maximal when all four states are at similar energies, and fall off rapidly with the energy difference between the states involved \cite{Mirlin, Gornyi, BAA, antomuller}. We therefore follow \cite{ Gornyi, BAA, antomuller} and restrict ourselves to considering processes where all states are at a similar energy $E_{\alpha} \approx E_{\beta} \approx E_{\gamma} \approx E_{\delta} \approx E$. In Appendix A we relax this approximation, and argue that this should yield qualitatively similar results. 

At a given energy $E$, the localization length is $\xi(E)$, given by (\ref{xi}). There are $\xi^{2} P(E)$ occupied states $\alpha$ that may scatter through the interaction at a particular point in space, and $\xi^2 P(E)$ occupied states $\beta$. Meanwhile, there are $\xi^{2}(1-P(E))$ Pauli-unblocked final states $\gamma$ and likewise for $\delta$. Thus, there are $\xi^8 [P(E)(1-P(E))]^2$ possible transitions that may be mediated by the interaction at a particular point in space, and thus the minimum energy change accessible through acting with the interaction at a particular typical point in space is $\sim \xi^{-8}[P(E)(1-P(E))]^2$. Meanwhile, the matrix element for the interaction is $ g \int d^2r \psi_{\alpha}^*(r) \psi_{\beta}^*(r) \psi_{\gamma}(r) \psi_{\delta}(r)$. The integrand is only non-zero over a volume of order $\xi^2$, and in this volume has magnitude $\xi^{-4}$ (this follows because all the states are normalized such that $\int d^2 r |\psi|^2 = 1$). As a result, the matrix element is of order $g \xi^{-2}$. 
The ratio $\zeta$ of matrix element to accessible level spacing is thus
\begin{eqnarray}
\zeta(E) &=& g \xi^6(E) [P(E) (1- P(E))]^2
\end{eqnarray}
We note in particular that $\zeta$ is the quantity that controls perturbative analysis of the form \cite{Fleishman, BAA}. If perturbation theory is to converge, then we require that $\zeta(E)<1$ for all energies $E$. In contrast, if $\zeta(E)$ exceeds one at some energies for even infinitesimal $g$, then there is no regime of perturbative stability of localization. 

Now, in the expression for $\zeta(E)$ above, we are free to take $g$ to be small. Moreover, $P(E)(1-P(E)) \le 1$. Thus, the only possible danger comes from the factor of $\xi(E)$, which diverges exponentially fast at large energies. We thus examine the behavior of $\zeta(E)$ in the limit $E\rightarrow \infty$. If it diverges, then perturbation theory will break down for arbitrarily small $g$, whereas if it remains bounded in the limit $E \rightarrow \infty$, then it will be possible to make perturbation theory converge by taking $g$ to be sufficiently small. In the limit of interest $E \rightarrow \infty$, the Fermi-Dirac occupation probabilities can be replaced by the Boltzman forms $P(E) \sim \frac{n}{k_BT} \exp(-E/k_BT) $ and $1-P(E) \sim 1$. It is now convenient to introduce a critical temperature $T_c$, defined as 
\begin{equation}
k_B T_c = \frac{m \pi W^2 a^2 }{3 \hbar^2} = \frac{\pi a^2}{6 \lambda_W^2} W
\end{equation}
The control parameter $\zeta$ in the $E \rightarrow \infty$ limit then becomes
\begin{equation}
\zeta(E) = g \frac{8\hbar^6  E^3}{(3\pi)^6 m^3 T_c^6} \exp\left[\frac{E}{k_B} \left(\frac{1}{T_c} - \frac{1}{T} \right) \right]
\end{equation}
clearly, for $T > T_c$, $\zeta(E)$ can be made arbitrarily large by going to large $E$, at any non-zero $g$. Thus, for $T>T_c$, localization is necessarily unstable to arbitrarily weak interactions. This is in sharp contrast to one dimension, where there is a regime of stability at any finite temperature (Fig.1). In contrast, for $T < T_c$, $\zeta(E)$ is bounded for $E \rightarrow \infty$, and thus perturbation theory converges at a typical point in space. For $T < T_c$, the ratio $\zeta$ is maximized by $E = 3 k_B T T_c/(T_c-T)$. Meanwhile the critical interaction strength for breakdown of perturbation theory at a typical point in space $g_c(T<T_c)$ is set by $\max(\zeta(E)) = 1$, and takes the form
\begin{equation}
g_c = \frac{(3 m T_c)^3 \pi^6 \exp(3)}{8 \hbar^6 k_B^3} \left(\frac{T_c-T}{T} \right)^3
\end{equation}
(this formula is accurate in the vicinity of $T_c$). As $T\rightarrow T_c$, $g_c \rightarrow 0$ as $(T_c-T)^3$. This is illustrated in Fig.1. The main difference to one dimension is that in 1D the function $g_c(T)$ is non-zero for any finite $T$. In contrast, in the 2D continuum, $g_c$ goes to zero at $T_c$ and at temperatures higher than $T_c$ there is no localized phase for arbitrarily weak interactions. If the initial condition is not a Gibbs state, then the requirement for perturbation theory to converge is that the density of particles at high energy $E$ must fall off faster than $\exp(-E/k_B T_c)$. 

Thus we have shown that there is a region in the interaction strength-temperature plane where perturbation theory converges at a typical point in space. We note that this does not necessarily mean the system is many body localized, since there could still be rare regions that contain high energy particles, where the energy density is locally above $T_c$. In such rare regions, perturbation theory does not converge (Appendix B). This problem is generic to systems with a many body mobility edge, including the system examined in Ref.\onlinecite{BAA}. If the rare regions are ultimately localized in space, then the problem will display true MBL. In contrast, if the rare regions are mobile, then they will ultimately allow the system to thermalize, although the thermalization timescale may be extremely long when these regions are exponentially dilute. This latter possibility is what was referred to as `asymptotic many body localization' in Ref.\onlinecite{deroeck} and as `quantum many body glass' in Ref.\onlinecite{glass}. Whether the two dimensional continuum system in the impurity disorder limit is truly MBL, or is `only' a quantum many body glass, requires a detailed discussion of these rare regions, which is beyond the scope of the present paper and will be presented elsewhere. However, for the moment we simply note that even if the rare regions are ultimately mobile, on short timescales (and in finite sized systems) their effect will not be apparent, such that the system will still appear many body localized. Thus for the remainder of this paper, we simply refer to the region enclosed by the curve $g_c(T)$ as MBL, bearing in mind that this is shorthand for `MBL, modulo a potential delocalization in thermodynamically large samples and on the longest timescales due to exponentially rare regions.'

We therefore conclude that (at least asymptotic) many body localization {\it can} occur in the two dimensional continuum, provided the temperature is below a non-zero critical temperature $T_c$, at least within the impurity model for disorder, where the Fourier transform of the disorder potential is taken to have constant weight at all wave vectors. Above $T_c$, {\it arbitrarily} weak interactions lead to a breakdown of perturbation theory, which we interpret as a breakdown of localization driven by interactions. 

\begin{figure}
\includegraphics[width = \columnwidth]{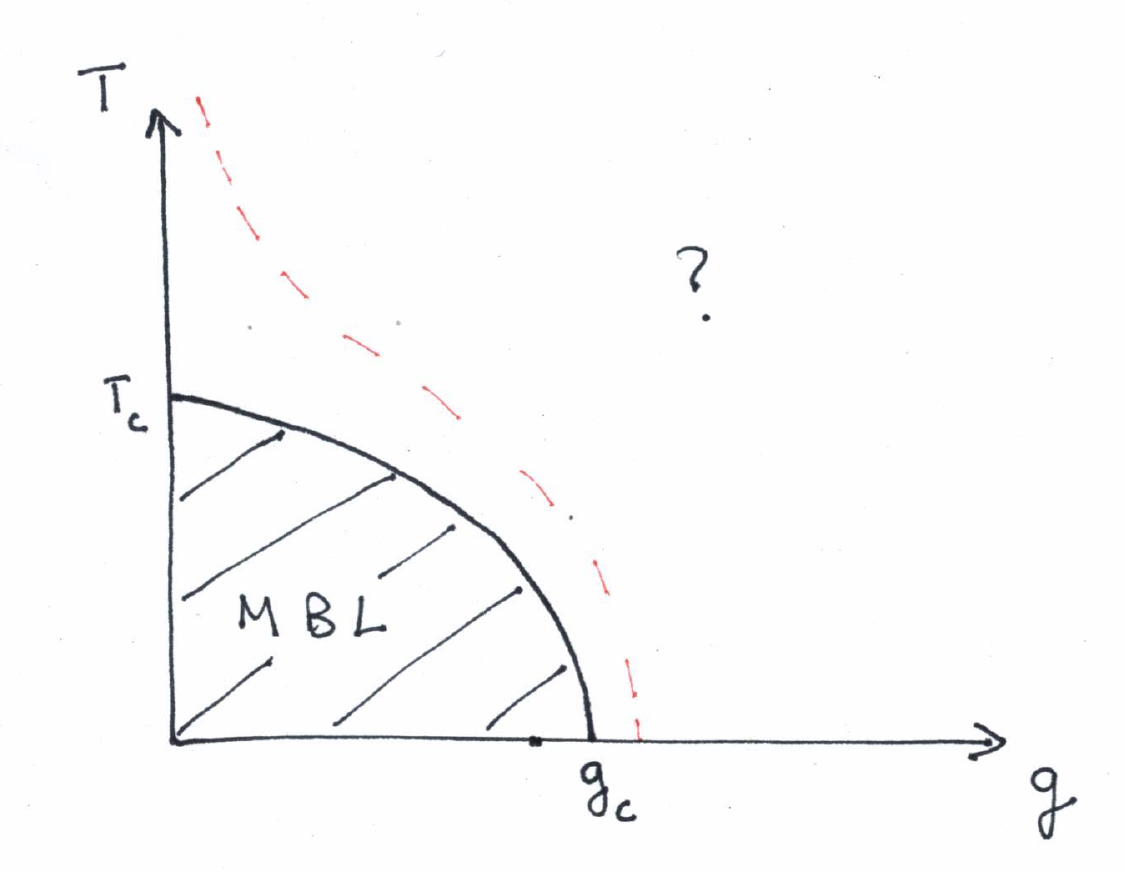}
\caption{Schematic phase diagram for the 2D continuum with impurity model disorder. Perturbation theory about the localized phase converges almost everywhere for temperatures $T < T_c$, indicating that (at least asymptotic) many body localization can occur at low enough temperature. For $T > T_c$, perturbation theory about the localized phase diverges for arbitrarily weak interactions. What happens in this regime cannot be answered within our current calculation scheme, but the breakdown of localization may represent interaction mediated delocalization. These results should be contrasted with one dimensional continuum systems (dashed line), where localization can be perturbatively stable at arbitrary temperature, at least if we ignore the potentially delocalizing effect of rare regions (Appendix B). \label{Fig1}}
\end{figure}

\subsection{ Marginal localization}
We have established that for $T<T_c$ there is a regime of MBL. However, even in this regime there is a `tail' of states with unboundedly large localization lengths. We will now demonstrate that the resulting many body localized phase exhibits `marginally localized' behavior, in the language of Ref.\onlinecite{RNACP}. 

We begin by noting that the relation between localization length and energy (10) may be inverted to yield
\begin{equation}
E(\xi) = \frac{m \pi W^2 a^2}{\hbar^2} \ln \frac{\hbar^2 2^{1/2} \xi}{m \pi^{3/2} W a}
\end{equation}
upto log log corrections. Now, the probability that a state with energy $E$ is occupied is $\frac{1}{k_BT}\exp(-E/k_BT)$. The probability that a state with localization length $\xi$ is occupied is (keeping track of Jacobian factors)
\begin{equation}
P(\xi) = \left(\frac{m \pi^{3/2} W a}{\hbar^2 w^{1/2} \xi}\right)^{3 T_c/T} \frac{3 T_c}{\xi T} \sim \xi^{-[1+3 T_c/T]}
\end{equation}
i.e. there is a power law tail to large $\xi$. Now let us examine the conductivity of a non-interacting system prepared with this distribution (the interacting system should behave qualitatively similarly, since the perturbation theory converges almost everywhere). The conductivity will be given by an expression of the form
\begin{equation}
\sigma \sim \int_{\xi L)}^{\infty} d\xi D(\xi) P(\xi) \nu(\xi)
\end{equation}
where $D$ is the diffusion constant, and $\nu$ is the density of states. We note that $D \sim \xi^{2-d}$ is independent of $\xi$ in two dimensions. We note also that the density of states is independent of energy (and hence independent of $\xi$) for quadratically dispersing particles in two dimensions. Thus, we have 
\begin{equation}
\sigma \sim D_0 \nu_0 \int_L^{\infty} d\xi \xi^{-(1+3 T_c/T)} \sim \frac{T}{T_c} L^{-3 T_c/T}
\end{equation}
i.e. the conductivity goes to zero in the thermodynamic limit, but only as a power law function of $L$. In this respect this system behaves as if it were `marginally localized' in the language of Ref.\onlinecite{RNACP}. 

\section{Beyond the impurity model}
Thus far we have worked with an `impurity' model of disorder, in which the Fourier transform of the two point correlation function of the disorder scattering potential is assumed to have equal weight at all wave vectors. However, this model is appropriate only when the disorder correlation length is asymptotically shorter than all other length scales in the problem. If the disorder correlation length is not asymptotically short, then the energy dependence of the two point correlation function must be taken into account. For almost all disorder potentials, $\gamma(E)$ will be a decreasing function of energy. In this case the localization length (6) will grow {\it faster} than exponentially with energy, overwhelming the suppressing effect of the $P(E)$ terms in (11). In this case, there will inevitably be a breakdown of localization at high energies, for arbitrarily weak interactions, and the stable regime of MBL will disappear. 

A particularly relevant (non-impurity) model of disorder is the speckle potential, which is widely used in cold atoms. With a speckle potential, the Fourier transform of the disorder correlation function abruptly cuts off  \cite{Palencia} for wave vectors $k > 1/a$, such that within Born approximation there is an apparent mobility edge at the wave vector $1/a$ (or equivalently at the temperature $T_m = \frac{\hbar^2 a^2}{2 m k_B}$). If we go beyond the Born approximation then all states will indeed be localized in the non-interacting limit, but the single particle localization length will jump to much larger values at the `apparent mobility edge.' Moreover, beyond the apparent mobility edge, the growth of the localization length with energy will be faster than exponential - and an exponential growth of single particle localization length with energy was already marginal for convergence of perturbation theory. 

In such a scenario there can be no true many body localization, since the handful of particles on the `delocalized' side of the apparent mobility edge can serve as a bath for the rest of the system. Of course, if the cutoff length scale is short enough, so that $T_c \ll T_m $, then the fact that we are dealing with a speckle rather than an impurity model disorder may not be of much practical relevance - the states above the apparent mobility edge will be occupied so (exponentially) rarely that the system will look apparently many body localized. In a finite sized system with $N$ particles, once $N \exp(-T_m / T_c) \ll 1$, then the high energy states that are the source of the trouble will almost surely be unoccupied, and should not affect the system. It is also possible that even though the perturbation theory inevitably breaks down (due to resonances at high energies) there is nevertheless a stable localized phase - albeit one that is not smoothly connected to the non-interacting phase - but this cannot be accessed through a perturbative calculation. 

Assuming that the breakdown of perturbation theory does indicate many body delocalization, there can be no true many body localized phase in the two dimensional continuum, away from the impurity limit on disorder. However, there can be a `nearly localized' regime. The `nearly localized' regime has been studied recently in a series of papers \cite{ngh, johri, narrowbath}. It has been argued that the `nearly localized' regime is well characterized by a `spectral line width' $\Gamma$, which can also be interpreted as the spin echo decay rate \cite{ngh}. As the localized phase is approached, this line width $\Gamma$ goes to zero. Proximity to localization for a two dimensional continuum system in a speckle potential may also be characterized in terms of its line width $\Gamma$. This line width will be small when $T / T_m \ll 1$ for two independent reasons. Firstly, the coupling between low energy states and the high energy `bath' will be weak \cite{Mirlin}, and will be suppressed by a polynomial function of $T_m-T$. When the coupling to the bath is weak, the spin echo decay rate is small. Indeed, for a system coupled to a bath with a coupling $g$, $\Gamma \sim g^2 \ln^d(1/g^2)$ (see Ref.\cite{ngh, johri}), where $d=2$ is the spatial dimensionality. Thus, as $T_m \rightarrow \infty$, the decoupling of high and low energy states will cause the line width to scale to zero as a power law function of $T_m$ (up to log corrections). Additionally, in the limit $T_m \rightarrow \infty$, there will be very few particles with energies above $k_B T_m$, and so the high energy `bath' responsible for delocalizing the system will be extremely dilute. The behavior of systems coupled to dilute baths will be discussed elsewhere. However, it is reasonable to expect that the diluteness of the bath will also contribute to the smallness of $\Gamma$ (and since the diluteness of the bath is an exponential function of $T_m$, this may actually be the dominant source of suppression of $\Gamma$). Certainly the line broadening $\Gamma$ (aka the spin echo decay rate), should vanish in the limit $T_m \rightarrow \infty$. Thus, in this limit, the delocalizing effect of the states above the apparent mobility edge becomes asymptotically weak, and there appears a regime (as in Fig.1) which is asymptotically close to localization. Still, at any finite $T_m$, with speckle potential disorder, the `phase boundary' drawn in Fig.1 cannot be shown to be a true phase boundary, but could be only a crossover. 

\section{Conclusions}

Thus, we conclude that in the disordered two dimensional continuum, localization can be perturbatively stable to interactions only if the disorder is in the impurity limit, and if the temperature is below a critical temperature (Fig.1). In this regime, the system exhibits sub diffusive relaxation, constituting marginal localization in the sense of Ref.\onlinecite{RNACP}. Whether rare region effects modify this conclusion is a subject of ongoing research, and is beyond the scope of the present paper. If the disorder is not in the impurity limit (e.g. disorder from a speckle potential), then there is no regime of perturbative stability for localization in the two dimensional continuum. However, in the limit where the disorder correlation length is short compared to all other length scales (i.e. in the impurity limit), the system can approach arbitrarily close to localization, in the sense of Ref.\onlinecite{ngh}. These ideas may be directly tested in cold atoms experiments. 

{\it Acknowledgements} I thank David A. Huse, S. S. Kondov, Waseem S. Bakr and Antonello Scardicchio for several stimulating conversations. I also thank Sarang Gopalakrishnan and Andrew C. Potter for feedback on the manuscript. I am supported by a PCTS fellowship.

\appendix

\section{Convergence of perturbation theory at a typical point in space}
In this appendix we consider processes whereby particles at very different energies interact. Of particular interest are processes whereby a high energy particle interacts with low energy particles. Let us first consider processes whereby a high energy particle with energy $E_{\alpha}$ scatters off a low energy particle with energy $E_{\beta} \ll E_{\alpha}$. At a given point in space, there are $\xi_{\alpha}^d$ states with energy $E_{\alpha}$ that have reasonable support, and $\xi_{\beta}^d$ states with energy $E_{\beta}$, and these are occupied with probabilities $P(E_{\alpha})$ and $P(E_{\beta})$ respectively. The number of possible initial states is thus $\xi_{\alpha}^d \xi_{\beta}^d P(E_{\alpha}) P(E_{\beta})$. These particles then scatter to final states with energies $E_{\gamma}$ and $E_{\delta}$, subject to the constraint $E_{\alpha} + E_{\beta} = E_{\gamma} + E_{\delta}$. There are $\xi_{\gamma}^d \xi_{\delta}^d (1-P(E_{\gamma})) (E-P(E_{\delta}))$ available final states, and so the number of possible transitions that can occur at a typical point in space is $\xi^d_{\alpha} \xi^d_{\beta} \xi^d_{\gamma} \xi^d_{\delta} P(E_{\alpha}) P(E_{\beta}) (1-P(E_{\gamma}))(1-P(E_{\delta})).$ The minimum energy change for a transition at a typical point in space is thus 
\begin{equation}
\delta E \sim \left[\xi^d_{\alpha} \xi^d_{\beta} \xi^d_{\gamma} \xi^d_{\delta} P(E_{\alpha}) P(E_{\beta}) (1-P(E_{\gamma}))(1-P(E_{\delta})) \right]^{-1}
\end{equation}
Meanwhile, if the interaction is contact, the matrix element is $\int g \psi^*_{\alpha} \psi^*_{\beta} \psi_{\gamma}\psi_{\delta}$, which is of order%
\begin{eqnarray}
M_{fi} &\sim& \frac{g}{ (E_{\alpha} - E_{\beta})^{\eta}} \frac{\min(\xi_{\alpha}^d, \xi^d_{\beta}, \xi^d_{\gamma}, \xi^d_{\delta})}{( \xi_{\alpha} \xi_{\beta} \xi_{\gamma} \xi_{\delta} )^{d/2}}
\end{eqnarray}
where $\eta$ is the power governing the falloff off the matrix element with the energy difference \cite{Mirlin}, and the factors of $\xi$ in the denominator come from normalization of the wave functions. Since the localization lengths $\xi$ are all exponential functions of the energy, the power law coefficient can be ignored, and the ratio $\zeta$ of matrix elements to energy change becomes 
\begin{widetext}
\begin{equation}
\zeta \sim g \min(\xi_{\alpha}^d, \xi^d_{\beta}, \xi^d_{\gamma}, \xi^d_{\delta}) ( \xi_{\alpha} \xi_{\beta} \xi_{\gamma} \xi_{\delta} )^{d/2}  P(E_{\alpha}) P(E_{\beta}) (1-P(E_{\gamma}))(1-P(E_{\delta})) 
\end{equation}
\end{widetext}
Now, when a high energy particle scatters off a low energy particle, if the two final states are at intermediate energies $E_{\gamma} \approx E_{\delta}$, then $\zeta$ is controlled by the initial high energy state $\alpha$, and is of order 
\begin{equation}
\zeta \sim g \xi_{\alpha} P(E_{\alpha}) 
\end{equation}
In this case, $\zeta$ is a decreasing function of $E_{\alpha}$ for low enough temperatures (including at all temperatures $T < T_c$), indicating convergence of perturbation theory.

We can also consider processes wherein $E_{\gamma} \approx E_{\alpha}$ and $E_{\delta} \approx E_{\beta}$ i.e. where a high and low energy particle scatter to a high and low energy state respectively. In this case, we have 
\begin{equation}
\zeta \sim g \xi_{\alpha}^{2} P(E_{\alpha})
\end{equation}
which again is a decreasing function of energy for low enough temperatures, including all temperatures $T < T_c$. Thus, consideration of processes whereby particles at very different energies interact does not alter the conclusion that (for impurity model disorder) perturbation theory converges almost everywhere at temperatures $T < T_c$, as asserted in the main text. 

\section{Rare region obstructions to localization?}

We note  that the convergence of perturbation theory at a typical point in space follows from the smallness of $P(E_{\alpha})$ - i.e. from the very low likelihood that a high energy particle is present at a typical point in space. Nevertheless, at some points in an infinite sample, there will be particles at arbitrarily high energies. We may wonder if these particles - where present - could serve as `nucleation centers for delocalization,' such that the resulting system is not truly MBL, but is rather a `quantum many body glass' (in the language of Ref.\onlinecite{glass}), where exponentially rare (but mobile) ergodic spots ultimately destroy localization.  

 Let us consider a situation where at a particular point in space, there is a particle with an arbitrarily large energy $E_{\alpha}$. This particle can interact with other particles in states at energy $E_{\beta}$. There are $\xi^{d}_{\alpha}$ states with energy $E_{\beta}$ with which this particle has significant overlap, and each of these is occupied with a probability $P(E_{\beta})$. The two particles can then scatter to final states with energy $E_{\gamma}$ and $E_{\delta}$ respectively, subject to the constraint $E_{\gamma} + E_{\delta} = E_{\alpha}+E_{\beta}$. Since the matrix element is only non-zero if all four states $\alpha, \beta, \gamma, \delta$ overlap, there are $\max(\xi_{\beta}^d, \xi_{\gamma}^d)$ choices for $\gamma$, and $\max(\xi^d_{\beta}, \xi^d_{\delta}) $ choices for $\delta$. 

When a high energy particle scatters off a low energy particle, one possibility is that the final states will be at an intermediate energy $E_{\beta} \ll E_{\gamma}, E_{\delta} \ll E_{\alpha}$. Iterated a few times, such processes will take the surplus energy from the particle $E_{\alpha}$ and will distribute it among a handful of particles, until all particles are at roughly the same energy, at which point the analysis in the main paper will apply. However, we wish to be conservative, and want to study all possibilities. Therefore, we consider the possibility where we scatter to final states with $E_{\gamma} \approx E_{\alpha}$ and $E_{\delta} \approx E_{\beta}$ (i.e. where a high and low energy initial state scatter to a high and low energy final state). 

In this case, the number of possible scattering processes will be $\xi^{2d}_{\alpha} \xi^d_{\beta} P(E_{\beta}) (1-P(E_{\beta}))$, and the accessible level spacing will be 
\begin{equation}
\delta E \sim \xi^{-2d}_{\alpha} \xi^{-d}_{\beta}
\end{equation}
noting that $P(E_{\beta})$ should not be small, since $\beta$ is a low energy state by postulate. 

The ratio of matrix element to accessible level spacing then becomes
\begin{eqnarray}
\zeta = \frac{M_{fi}}{\delta E} &\sim& g \xi_{\alpha}^d \xi_{\beta}^d
\end{eqnarray}
Since small $\zeta$ controls the perturbation theory, and since $E_{\alpha}$ (and hence $\xi_{\alpha}$ can be arbitrarily large, this (at least naively) presents a problem for perturbation theory. Could it be that exponentially rare high energy particles could repeatedly undergo resonant scatterings of low energy particles without being scattered down to low energies, and thus could end up diffusing over the system? And could this handful of high energy particles then act as a bath for the rest of the system, allowing it to thermalize on the longest timescales, such that the two dimensional continuum problem is not truly many body localized, but is only a quantum many body glass \cite{withanto}? A related question involves rare regions where the energy density is much higher than average, such that these regions appear to be at a temperature above $T_c$. Could such regions serve as delocalization centers, allowing the system to thermalize on the longest timescales? These questions are beyond the scope of the present paper, and will be discussed elsewhere \cite{withanto}. We note however that even if such `delocalization' mediated by exponentially rare high energy particles does occur, it will become apparent only on enormously large timescales, such that on experimentally relevant timescales the system could be indistinguishable from a truly localized state \cite{glass}.


\begin{thebibliography}{99}
\bibitem[Anderson (1958)]{pwa}
P. W. Anderson, Phys. Rev. {\bf 109}, 1492 (1958).
\bibitem[LeeRMP (1980)]{LeeRMP}
P.A. Lee and T.V. Ramakrishnan, Rev. Mod. Phys. 57, 287 (1985)
\bibitem{Fleishman}
L. Fleishman and P. W. Anderson, Phys. Rev. B 21, 2366 (1980).
\bibitem[AKGL (1997)]{agkl}
B. L. Altshuler, Y. Gefen, A. Kamenev and L. S. Levitov, Phys. Rev. Lett. {\bf 78}, 2803 (1997).
\bibitem[Mirlin (2006)]{Gornyi}
I. V. Gornyi, A. D. Mirlin and D. G. Polyakov, Phys. Rev. Lett. {\bf 95}, 206603 (2005).
\bibitem[BAA (2006)]{BAA}
D. M. Basko, I. L. Aleiner and B. L. Altshuler, Annals of Physics {\bf 321}, 1126 (2006).
\bibitem{imbrie}
J. Z. Imbrie, arXiv:1403.7837.
\bibitem[Oganesyan (2008)]{Oganesyan}
V. Oganesyan and D. A. Huse,
Phys. Rev. B {\bf 75}, 155111 (2007).
\bibitem[Znid (2008)]{Znid}
M. Znidaric, T. Prosen and P. Prelovsek, Phys. Rev. B {\bf 77}, 064426 (2008).
\bibitem[Pal (2010)]{pal}
A. Pal and D. A. Huse,
Phys. Rev. B {\bf 82}, 174411 (2010).
\bibitem{arcmp}
R. Nandkishore and D. A. Huse, arXiv:1404.0686.
\bibitem[Lbits (2013)]{Lbits}
D. A. Huse and V. Oganesyan, arXiv:1305.4915; D. A. Huse, R. Nandkishore and V. Oganesyan arXiv: 1408.4297
\bibitem[Abanin (2013)]{Abanin}
M. Serbyn, Z. Papic and D. A. Abanin, Phys. Rev. Lett. {\bf 111}, 127201 (2013).
\bibitem[Deutsch (1991)]{Deutsch}
J. M. Deutsch, Phys. Rev. A {\bf 43}, 2046 (1991).
\bibitem[Srednicki (1994)]{Srednicki}
M. Srednicki, Phys. Rev. E {\bf 50}, 888 (1994).
\bibitem[Rigol (2008)]{Rigol}
M. Rigol, V. Dunjko and M. Olshanii, Nature {\bf 452}, 854 (2008).
\bibitem[ngh (2014)]{ngh}
R. Nandkishore, S. Gopalakrishnan and D.A. Huse, arXiv: 1402.5971
\bibitem[johri (2014)]{johri}
S. Johri, R. Nandkishore and R.N. Bhatt, arXiv: 1405.5515 (2014)
\bibitem[LPQO (2013)]{LPQO}
D. A. Huse, R. Nandkishore, V. Oganesyan, A. Pal and S. L. Sondhi, Phys. Rev. B {\bf 88}, 014206 (2013).
\bibitem[Pekker (2013)]{Pekker}
D. Pekker, G. Refael, E. Altman, E. Demler and V. Oganesyan, Phys. Rev. X {\bf 4}, 011052 (2014).
\bibitem[Vosk (2013)]{Vosk}
R. Vosk and E. Altman, Phys. Rev. Lett. {\bf 112}, 217204 (2014).
\bibitem[Bauer (2013)]{Bauer}
B. Bauer and C. Nayak, J. Stat. Mech. P09005 (2013).
\bibitem[Bahri (2013)]{Bahri}
Y. Bahri, R. Vosk, E. Altman and A. Vishwanath, arXiv:1307.4192.
\bibitem{lspt} A. Chandran, V. Khemani, C. R. Laumann and S. L. Sondhi, Phys. Rev. B {\bf 89}, 144201 (2014).
\bibitem[grover (2013)]{grover}T. Grover and M. P. A. Fisher, arXiv:1307.2288.
\bibitem[Muller (2013)]{Muller} M. Schiulaz, M. Muller, arXiv:1309.1082.
\bibitem[deroeck (2013)]{deroeck} F. Huveneers and W. De Roeck, arXiv:1405.3279.
\bibitem[glass (2015)]{glass} R. Nandkishore, D.A. Huse and S.L. Sondhi, in preparation. 
\bibitem[antomuller (2014)]{antomuller}
V. Ros, M. Muller and A. Scardicchio, arXiv: 1406.2175
\bibitem[RNACP (2014)]{RNACP}
R. Nandkishore and A.C. Potter, arXiv: 1406.0847
\bibitem[Aleiner (2008)]{Aleiner}
I.L. Aleiner, B.L. Altshuler and G.V. Shlyapnikov, {\it Nature Physics} {\bf 6}, 900-904 (2010)
\bibitem[Narrowbath (2014)]{narrowbath}
S. Gopalakrishnan and R. Nandkishore, arXiv: 1405.1036
\bibitem[Mirlin (2000)]{Mirlin}
A.D. Mirlin, {\it Phys. Rep. } {\bf 326}, 259-382 (2000)
\bibitem[Palencia (2008)]{Palencia} L. Sanhez-Palencia,  D Clement, P Lugan, P Bouyer and A Aspect, New J. Phys. {\bf 10} 045019 (2008)
\bibitem[withanto (2014)]{withanto}
R. Nandkishore {\it et al}, in preparation

\end{thebibliography}
\end{document}